\documentclass[draftcls, onecolumn, journal, 12pt, letterpaper]{IEEEtran}
\usepackage{latexsym}
\usepackage{amsmath}
\usepackage{amsfonts}
\usepackage{rawfonts,amsthm}
\usepackage{amssymb}
\usepackage{float}
\usepackage{paralist}
\usepackage{verbatim}
\usepackage{mathrsfs}
\usepackage{multirow}
\usepackage{amsfonts}
\usepackage{longtable}
\usepackage{mathrsfs}
\usepackage{ifthen}
\usepackage{listings}
\usepackage{url}
\usepackage{extarrows}

\usepackage{graphicx,color}
\usepackage{subfigure}
\usepackage{xcolor}
\usepackage[framemethod=TikZ]{mdframed}
\usepackage{framed}
\usepackage{mathrsfs}
\usepackage{multicol}

\usepackage{calligra}
\usetikzlibrary{arrows}
\usetikzlibrary{decorations.pathreplacing}
\usepackage{threeparttable}
\usepackage{graphicx}
\usepackage{epstopdf}
\usepackage{epsfig}
\usepackage{algpseudocode}
\usepackage{soul} 

\begin{document}

\title{Decoding Polar Codes via Weighted-Window Soft Cancellation for Slowly-Varying Channel}

\author{
	Yong~Fang
	\thanks{The author is with the School of Information Engineering, Chang'an University, Xi'an, Shaanxi 710064, China (email: fy@chd.edu.cn).}
}


\maketitle

\begin{abstract}
	Polar codes are a class of {\bf structured} channel codes proposed by Ar{\i}kan based on the principle of {\bf channel polarization}, and can {\bf achieve} the symmetric capacity of any Binary-input Discrete Memoryless Channel (B-DMC). The Soft CANcellation (SCAN) is a {\bf low-complexity} {\bf iterative} decoding algorithm of polar codes outperforming the widely-used Successive Cancellation (SC). Currently, in most cases, it is assumed that channel state is perfectly {\bf known} at the decoder and remains {\bf constant} during each codeword, which, however, is usually unrealistic. To decode polar codes for {\bf slowly-varying} channel with {\bf unknown} state, on the basis of SCAN, we propose the Weighted-Window SCAN (W$^2$SCAN). Initially, the decoder is seeded with a coarse estimate of channel state. Then after {\bf each} SCAN iteration, the decoder progressively refines the estimate of channel state with the {\bf quadratic programming}. The experimental results prove the significant superiority of W$^2$SCAN to SCAN and SC. In addition, a simple method is proposed to verify the correctness of SCAN decoding which requires neither Cyclic Redundancy Check (CRC) checksum nor Hash digest.
\end{abstract}

\begin{IEEEkeywords}
	Polar codes, Slowly-varying channel, Soft cancellation, Channel estimation.%
\end{IEEEkeywords}

\newpage

\section{Introduction}
\IEEEPARstart{C}{hannel} coding is an old issue whose theoretical foundations were laid by C.~Shannon in his seminal papers \cite{ShannonA,ShannonB}. The first class of modern channel codes are Hamming codes \cite{Hamming}. Since then, many kinds of good channel codes are designed, e.g., Reed-Muller (RM) codes \cite{RMcodesR,RMcodesM}, Reed-Solomon (RS) codes \cite{RScodes}, and Bose-Chaudhuri-Hocquenghem (BCH) codes \cite{BCHcodesH,BCHcodesBC}. These channel codes can be classified into {\bf structured} codes due to their regular structures. From 1990s, most researchers turned their attention to {\bf random} codes, e.g., turbo codes \cite{turbo} and Low-Density Parity-Check (LDPC) codes \cite{LDPC}. Due to their excellent performance, random codes have found wide uses in many practical scenarios, e.g., turbo codes are used in 3G/4G mobile communications and satellite communications, while LDPC codes are adopted by the DVB-S2, ITU-T G.hn, 10GBase-T Ethernet, and Wi-Fi 802.11. 

In 2009, Ar{\i}kan proposed a very different idea for channel coding \cite{Arikan} based on a very simple finding: After an eXclusive OR (XOR) operation, two {\em independent} and {\em identical} {\bf physical} channels can form two {\em different} {\bf virtual} channels---a degraded virtual channel and an upgraded virtual channel. Let $W_1$ and $W_2$ be two physical channels and from them, two virtual channels $V_1$ and $V_2$ are formed. If $W_1$ and $W_2$ are mutually independent and $W_1=W_2=W$, where $W$ is a symmetric Binary-input Discrete Memoryless Channel (B-DMC), then $I(V_1)<I(V_2)$ and $I(V_1)+I(V_2)=2I(W)$, where $I(\cdot)$ denotes the capacity of a channel. By repeating this operation, $N=2^n$ independent and identical physical channels $W_i$'s can form $N$ different virtual channels $V_i$'s and $\sum_{i=1}^{N}{I(V_i)} = NI(W)$. Surprisingly, it is proved in \cite{Arikan} that, as $N\to\infty$, degraded virtual channels will become useless, i.e., their capacities tend to $0$; while upgraded virtual channels will become perfect, i.e., their capacities tend to $1$. More important, the fraction of perfect virtual channels will approach to $I(W)$, while the fraction of useless virtual channels will approach to $1-I(W)$ \cite{Arikan}. This phenomenon is called {\bf channel polarization}. Thus, we can sort all virtual channels and use $K<NI(W)$ best virtual channels to transmit user bits, while leaving the other $(N-K)$ virtual channels idle. Given code rate $R=K/N<I(W)$, if polar codes are decoded with the Successive Cancellation (SC), whose complexity is $O(N\log_2N)$, then the frame error probability $P_e(N,R)\leq 2^{-N^\beta}$ for any $\beta<0.5$ \cite{polarization}. 

As the first class of {\bf capacity-achievable} channel codes, polar codes have aroused keen interest from both academia and industry. People are working mainly on the following issues (Some of them have been solved or partially solved, while the others still remain open):
\begin{itemize}
	\item {\bf Systematic polar codes}. Polar codes were originally proposed in its asystematic form \cite{Arikan}. Later, Ar{\i}kan gave the systematic form of polar codes in \cite{syst}. A simplified encoding method for systematic polar codes was proposed in \cite{systenc}. 
	\item {\bf Capacities of virtual channels for general physical channels}. In the original paper \cite{Arikan}, only for Binary Erasure Channel (BEC), a recursive formula is given, while for general B-DMC, e.g., Binary Symmetric Channel (BSC) and Additive White Gaussian Noise (AWGN) channel, there is no efficient algorithm for the  capacity of virtual channel.
	\item {\bf Decoding algorithms}. Originally, Ar{\i}kan gave two decoding algorithms for polar codes: the SC and the Belief Propagation (BP) \cite{BP}. The SC is a {\bf non-iterative} algorithm with the {\bf lowest complexity}, but its efficiency is less than satisfactory, so the list decoder was used to improve its efficiency \cite{SCL}. Further, aided by Cyclic Redundancy Check (CRC), polar codes with the SC list decoder can even defeat LDPC codes \cite{crc}. On the contrary, the BP is an {\bf iterative} algorithm with the {\bf best efficiency}, but its complexity is very high, so it was simplified by the Soft CANcellation (SCAN) \cite{SCAN}, a {\bf low-complexity} {\bf iterative} algorithm that soundly balances efficiency and complexity.
	\item {\bf Extensions to general channel models}. Though polar codes were originally proposed for binary-input channels, they can be extended to nonbinary-input channels \cite{nonbinary}. 
	\item {\bf Extensions to general coding problems}. As other channel codes, polar codes can also be applied to different coding problems. Immediately after \cite{polarization}, Ar{\i}kan studied the mirror problem of channel polarization---source polarization in \cite{srcpolar}, which lays the theoretical foundation for the applications of polar codes to source coding. After that, polar codes have found wide uses in many scenarios, e.g., lossless/lossy source coding \cite{losslesssource,lossysource}, Joint Source-Channel Coding (JSCC) \cite{jscc}. In \cite{source}, polar codes are evaluated for two forms of Distributed Source Coding (DSC), i.e., Slepian-Wolf coding (lossless DSC) and Wyner-Ziv coding (asymmetric lossy DSC with decoder side information).
\end{itemize}

As pointed out in \cite{Arikan}, to construct polar codes best fit to the channel, we need three steps: (1) Calculating the capacities of virtual channels according to physical channels; (2) Sorting virtual channels according to their capacities; (3) Designating best virtual channels for user bits. Thus, a prerequisite for constructing optimal polar codes is that the exact states of physical channels are known in advance at both encoder and decoder. For stationary channels, this prerequisite is satisfiable because channel states can be estimated at the decoder by many methods, e.g., inserting additional pilot symbols into codewords \cite{chest}. However, for nonstationary channels, it is hardly possible to exactly trace the local state of the channel at each instant before decoding, making it impossible to design polar codes best fit to the channel before transmission. 

In this paper, we consider such a channel model---A symmetric B-DMC with {\bf unknown slowly-varying} state. It is proved in \cite{proof_ns_ml} that Ar{\i}kan's construction also polarizes {\bf time-varying} {\bf memoryless} channels in the same way as it polarizes {\bf stationary} {\bf memoryless} channels, which lays a theoretical foundation for the applications of polar codes to time-varying channels. Further in \cite{fastpolar}, to speedup the polarization of time-varying memoryless channels, Ar$\i$kan's channel polarization transformation is combined with certain permutations and skips at each polarization level. However in \cite{proof_ns_ml} and \cite{fastpolar}, local states of time-varying memoryless channels are assumed to be completely known at both encoder and decoder. 

In our prior papers \cite{FangTCOM12,FangTCOM13}, the problem of channel coding with unknown slowly-varying state has been extensively studied for LDPC codes. A variant of the BP algorithm, i.e., the so-called Sliding-Window BP (SWBP), was developed, which can exactly estimate the local state of slowly-varying channel at each instant during LDPC decoding. The SWBP is a pilot-free method for joint {\bf channel decoding} and {\bf state estimation} that possesses many merits, e.g, low complexity, high efficiency, and strong robustness. So naturally, we raise the following questions: Can the SWBP be extended to polar codes, and if yes, how to extend it to polar codes and how well it performs for polar codes?

The above questions will be answered by this paper. We show that on the basis of SCAN decoding, the core idea of SWBP can be applied to polar codes for joint channel decoding and state estimation.  The reason why we choose the SCAN as the platform to extend the SWBP from LDPC codes to polar codes is because the SCAN is an {\bf iterative} decoding algorithm with {\bf low complexity}. The main novelty of this paper is proposing the Weighted-Window SCAN (W$^2$SCAN) algorithm, which optimizes tap weights of sliding window by the quadratic programming. Another trivial novelty is proposing a simple method to verify the correctness of SCAN decoding, which needs neither CRC checksum nor Hash digest.

The rest of this paper is arranged as below. Section \ref{sec:model} defines a piecewise-stationary channel model for performance evaluation. Section \ref{sec:scan} describes the SCAN algorithm in detail, and also proposes a simple method to decide whether the decoding is successful or not in the absence of CRC checksums and Hash digests. Section \ref{sec:w2scan} deduces the W$^2$SCAN algorithm. Section \ref{sec:exp} reports simulation results. Finally, Section \ref{sec:con} concludes this paper.

\section{Piecewise-Stationary Channel Model}\label{sec:model}
Consider a time-varying channel $W: \mathcal{X}^N\times\mathcal{S}^N\to\mathcal{Y}^N$, where $\mathcal{X}$, $\mathcal{Y}$, and $\mathcal{S}$ are input space, output space, and state space of the channel, respectively. This paper considers only B-DMC, i.e., $\mathcal{X}=\mathbb{B}$. Let $x^N\triangleq(x_1,\cdots,x_N)\in\mathbb{B}^N$. The channel transition matrix is
\begin{align}
	W(y^N|x^N) = \sum_{s^N\in\mathcal{S}^N}p(s^N)W(y^N|x^N,s^N),
\end{align}
where $p(s^N) = p(s_1)\prod_{i=2}^{N}p(s_i|s^{i-1})$. Especially, if channel state information is memoryless, $p(s^N) = \prod_{i=1}^{N}p(s_i)$. Assume that the channel is {\bf conditionally-memoryless given state}, i.e.,
\begin{align}
	W(y^N|x^N,s^N) = \prod_{i=1}^{N}{W(y_i|x_i,s_i)}.
\end{align}
Let $\hat{C}$ be the channel capacity when state information is available at the decoder. Then \cite{Gamal12}
\begin{align}
	\hat{C} = \frac{1}{N}\max_{p(x^N)}I(X^N;Y^N|S^N) = \frac{1}{N}\sum_{i=1}^N \hat{C}_i,
\end{align}
where $\hat{C}_i=\max_{p(x_i)}I(X_i;Y_i|S_i)$. Let $C$ be the channel capacity when state information is available at neither encoder nor decoder. Then \cite{Gamal12}
\begin{align}
	C = \frac{1}{N}\max_{p(x^N)}I(X^N;Y^N).
\end{align}
It is prove that $C\leq \hat{C}$, and the equality holds if $S^N$ is constant \cite{Gamal12}.

Now we focus on varying state $S^N$, which may be the sequence of local crossover probabilities for time-varying BSC or the sequence of local noise variances for time-varying AWGN channel. Unfortunately, {\bf arbitrarily-varying} channel is intractable because it is impossible to estimate instantaneous state by samples. Thus for arbitrarily-varying channel, $C<\hat{C}$. On the contrary, if we impose a {\bf slowly-varying} constraint on the channel, it is possible to estimate instantaneous state by a number of {\bf consecutive} samples at the decoder. We can imagine this process as moving a {\bf sliding window} forwards and the only thorny point is how to set the size of sliding window. Then it looks like that state information of the slowly-varying channel is available at the decoder, so $C\approx \hat{C}$. In \cite{FangTCOM12} and \cite{FangTCOM13}, we assume that channel state $S^N$ varies sinusoidally, which is a too idealistic model. Below, we will define a more practical model, i.e., the piecewise-stationary channel, which is based on two assumptions:
\begin{itemize}
	\item Channel state remains constant during each piece, and the length of pieces obeys the $\lambda$-Poisson distribution. Let $K$ be the length of a piece. Then $\Pr(K=k) = \frac{e^{-\lambda}\lambda^k}{k!}$. 
	\item The state of each piece is a realization of random variable $S$. Let $s_j$ be the state of the $j$-th piece. Then $s_j$'s are independently and identically drawn from space $\mathcal{S}$.
\end{itemize}

Let $\hat{C}(s) \triangleq \max_{p(x)}I(X;Y|s)$, where $s\in\mathcal{S}$. Given state $S$ available at the decoder, channel capacity is $\hat{C} = \sum_{s\in\mathcal{S}}p(s)\hat{C}(s)$ or $\hat{C} = \int_{s\in\mathcal{S}}f(s)\hat{C}(s)ds$, where $p(\cdot)$ is the distribution of a discrete random variable and $f(\cdot)$ is the distribution density of a continuous random variable. The detailed form of $\hat{C}(s)$ depends on the used channel model. Following are two examples:
\begin{itemize}
	\item Let $\epsilon(s)$ be the crossover probability of a BSC given state $s\in\mathcal{S}$. Then $\hat{C}(s) = 1-H(\epsilon(s))$, where $H(\cdot)$ is the binary entropy function. 
	\item Let $\sigma^2(s)$ be the noise variance of an AWGN channel given state $s\in\mathcal{S}$. For the Binary Phase Shift Keying (BPSK) modulation, i.e., $0\to +1$ and $1\to -1$, we have $\hat{C}(s) = 1-\int_{-\infty}^{\infty}f(y|s)H(\epsilon(y|s))dy$, where
	\begin{align}\label{eq:epsilonys}
	\epsilon(y|s) \triangleq \frac{\exp(y/\sigma^2(s))}{\exp(-y/\sigma^2(s))+\exp(y/\sigma^2(s))}
	\end{align}
	and
	\begin{align}\label{eq:fys}
	f(y|s) \triangleq \frac{1}{2\sqrt{2\pi\sigma^2(s)}}\left(\exp(-\frac{(y-1)^2}{2\sigma^2(s)}) + \exp(-\frac{(y+1)^2}{2\sigma^2(s)})\right).
	\end{align}	
\end{itemize}
For a large $\lambda$, the piecewise-stationary channel is {\bf slowly-varying} and thus $C\approx\hat{C}$. 

If the decoder disregards the non-stationarity of the channel, a time-varying channel will look more or less like an equivalent stationary channel. For the BSC model, the constant crossover probability of the equivalent stationary channel is $\bar{\epsilon} = \int_{s\in\mathcal{S}}f(s)\epsilon(s)ds$ or $\bar{\epsilon} = \sum_{s\in\mathcal{S}}p(s)\epsilon(s)$, and its capacity is $\bar{C} = 1-H(\bar{\epsilon})$. For the AWGN model, the constant noise variance of the equivalent stationary channel is $\bar{\sigma}^2 = \int_{s\in\mathcal{S}}f(s)\sigma^2(s)ds$ or $\bar{\sigma}^2 = \sum_{s\in\mathcal{S}}p(s)\sigma^2(s)$, and its capacity is $\bar{C} = 1-\int_{-\infty}^{\infty}\bar{f}(y)H(\bar{\epsilon}(y))dy$, where $\bar{\epsilon}(y)$ and $\bar{f}(y)$ are obtained from \eqref{eq:epsilonys} and \eqref{eq:fys}, respectively, by replacing $\sigma^2(s)$ with $\bar{\sigma}^2$. Due to the concavity of entropy, $\bar{C}\leq \hat{C}$ and the equality holds {\em if and only if} (iff) the channel is indeed stationary. 

The above analysis shows that, for a {\bf slowly-varying} channel with {\bf unknown} state, instead of brutally treating the channel as a {\bf stationary} channel, a big gain $(C-\bar{C})\approx(\hat{C}-\bar{C})$ may be achieved if only the varying state of the channel at each instant is ingeniously estimated at the decoder. This potential gain motivates \cite{FangTCOM12, FangTCOM13}, and this paper to fully exploit the non-stationarity of slowly-varying channel.

\section{An Overview on Polar Codes}\label{sec:scan}
\begin{figure*}[!t]
	\small
	\centering
	\setlength{\unitlength}{1cm}
	\begin{tikzpicture}
	\foreach \y in {1,3,5,7} \node[circle,fill=green,label=$V_{3,\y}$,label=left:$u_{\y}$]at(0, 12-1.5*\y){+};   
	\foreach \y in {2,4,6,8} \node[circle,fill=green,label=$V_{3,\y}$,label=left:$u_{\y}$]at(0, 12-1.5*\y){=};
	\foreach \y in {1,2,5,6} \node[circle,fill=cyan,label=$V_{2,\y}$]at(4, 12-1.5*\y){+};   
	\foreach \y in {3,4,7,8} \node[circle,fill=cyan,label=$V_{2,\y}$]at(4, 12-1.5*\y){=};
	\foreach \y in {1,2,3,4} \node[circle,fill=red,label=$V_{1,\y}$]at(8, 12-1.5*\y){+};   
	\foreach \y in {5,6,7,8} \node[circle,fill=red,label=$V_{1,\y}$]at(8, 12-1.5*\y){=}; 	  
	\foreach \y in {1,...,8} \node[circle,fill=yellow,label=\qquad$V_{0,\y}/W_{\pi(\y)}$,label=right:$x_{\y}$]at(12, 12-1.5*\y){=};	 
	
	\foreach \y in {0,...,7} \draw[<-] (0.3,1.5*\y)--(4-0.3,1.5*\y);
	\foreach \y in {0,2,4,6} \draw[<-] (0.3,1.5*\y)--(4-0.3,1.5*\y+1.5);
	\foreach \y in {1,3,5,7} \draw[<-] (0.3,1.5*\y)--(4-0.3,1.5*\y-1.5);
	
	\foreach \y in {0,...,7} \draw[<-] (4.3,1.5*\y)--(8-0.3,1.5*\y);
	\foreach \y in {0,1,4,5} \draw[<-] (4.3,1.5*\y)--(8-0.3,1.5*\y+3);
	\foreach \y in {2,3,6,7} \draw[<-] (4.3,1.5*\y)--(8-0.3,1.5*\y-3);
	
	\foreach \y in {0,...,7} \draw[<-] (8.3,1.5*\y)--(12-0.3,1.5*\y);
	\foreach \y in {0,...,3} \draw[<-] (8.3,1.5*\y)--(12-0.3,1.5*\y+6);
	\foreach \y in {4,...,7} \draw[<-] (8.3,1.5*\y)--(12-0.3,1.5*\y-6);	
	\end{tikzpicture}
	\caption{An example of polar encoding and virtual channel construction for slowly-varying physical channel, where $x^N$ is the codeword of the message $u^N$ and $V_{l,i}$ is the $i$-th virtual sub-channel after $l$ levels of polarization. Note that before transmission, $x^N$ is permuted to decorrelate adjacent physical sub-channels with memory. The course of polarization is from right to left. Initially, $V_{0,i}=W_{\pi(i)}$, where $(\pi(1),\cdots,\pi(N))$ is a permutation of $(1,\cdots,N)$ and $W_i$ is the $i$-th physical sub-channel. The operation of channel degradation is denoted by $\textcircled{+}$, and the operation of channel upgradation is denoted by $\textcircled{=}$.}
	\label{fig:channel}
\end{figure*}
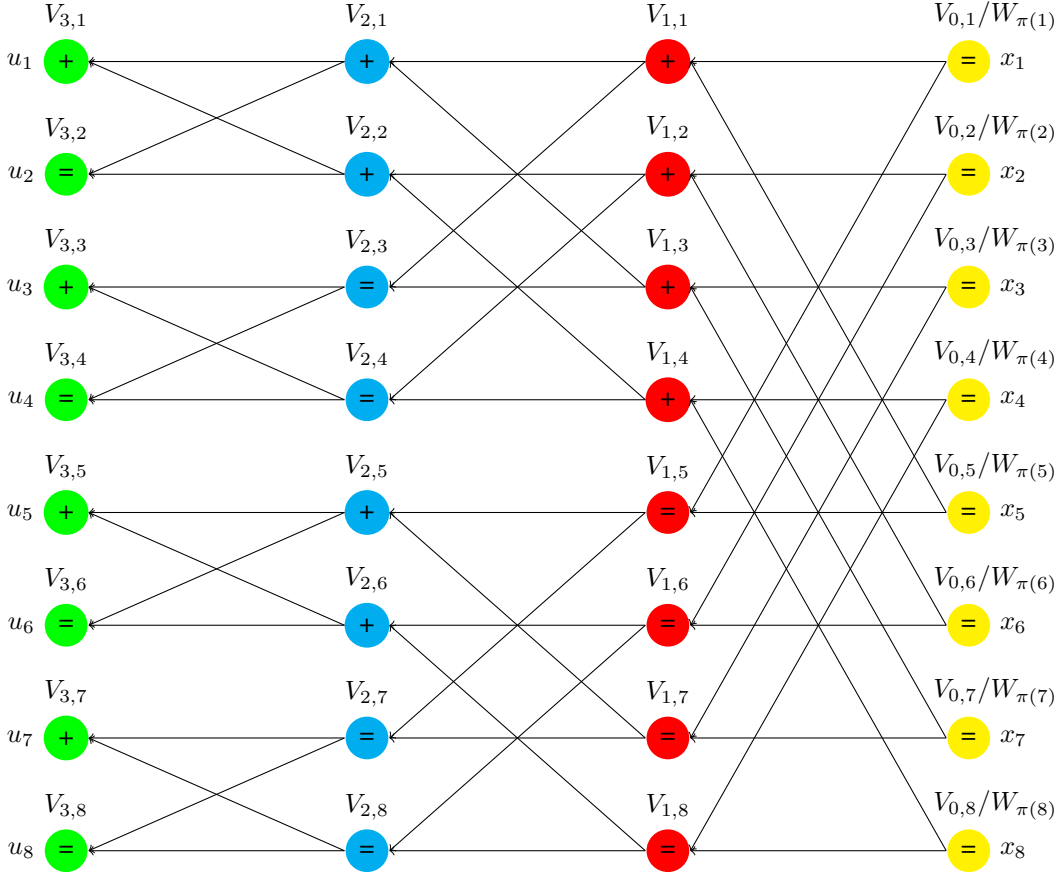

To develop the W$^2$SCAN algorithm, the reader must be equipped with some necessary basic knowledge of polar codes. In this section, we will first skim over the encoding of polar codes. Then we will describe the SCAN decoding of polar codes in detail, which is extremely important for the reader to understand the W$^2$SCAN algorithm that will be proposed in the next section. Finally, we will propose a simple method to decide the correctness of SCAN decoding without the aid of CRC checksums and Hash digests. Note that throughout the rest of this paper, $n=\log_2N\in\mathbb{Z}$ without explicit declaration, where $N$ is code length.

\subsection{Encoding of Polar Codes}
As shown by Fig.~\ref{fig:channel}, the message $u^N\in\mathbb{B}^N$ is encoded to get codeword $x^N = u^NG_N= u^N B_N F^{\otimes n}$, where $B_N$ permutes $u^N$ to be in the bit-reversed order, $F = (\begin{smallmatrix}1,0\\1,1\end{smallmatrix})$, and $\otimes$ denotes the Kronecker product. Then $x^N$ is transmitted over a noisy physical channel $W^N$. If $W^N$ is {\bf slowly-varying}, there will exist {\bf memory} between adjacent physical sub-channels, which prevents fast polarization of virtual sub-channels. To accelerate the polarization of virtual sub-channels, we decorrelate adjacent physical sub-channels by permuting $x^N$ before transmission. After permutation, virtual sub-channels can be quickly polarized. This is one purpose of the permutation step (another purpose will be given in Sect.~IV-B).

Fig.~\ref{fig:channel} shows how $N$ virtual sub-channels are constructed from $N$ physical sub-channels by $n$ levels of polarization. Let $V_{l,i}$ denote the $i$-th virtual sub-channel after $l$ levels of polarization, where $0\leq l\leq n$ and $1\leq i\leq N$. Initially, we set $V_{0,i}=W_{\pi(i)}$, where $(\pi(1),\cdots,\pi(N))$ is a permutation of $(1,\cdots,N)$. Then as $l$ increases from 1 to $n$, we have $(V_{l,i}, V_{l,i+2^{n-l}}) \leftarrow (V_{l-1,i}, V_{l-1,i+2^{n-l}})$, where $1\leq (i-k2^{n-l+1})\leq 2^{n-l}$ for $0 \leq k < 2^{l-1}$. Let $\textcircled{+}$ denote the operation of channel degradation and $\textcircled{=}$ denote the operation of channel upgradation. Then $V_{l,i} = V_{l-1,i} \textcircled{+} V_{l-1,i+2^{n-l}}$ and $V_{l,i+2^{n-l}} = V_{l-1,i} \textcircled{=} V_{l-1,i+2^{n-l}}$. Finally, $N$ virtual sub-channels $V_{n,i}$'s after $n$ levels of polarization are sorted according to their capacities. For an $(N,K)$ polar code, only $K$ virtual sub-channels with the largest capacities are occupied by user bits, while others are left idle. Let $\mathcal{A}$ be the set of indices of $K$ user virtual sub-channels and $\mathcal{A}_c = \{1,\cdots,N\}\setminus \mathcal{A}$. For $i\in\mathcal{A}_c$, $u_i$ is called a frozen bit and set to 0 usually \cite{Arikan}.

\subsection{Soft CANcellation (SCAN) Decoding}
Compared with LDPC codes, an advantage of polar codes is that they can be decoded at very low complexity by the SC \cite{Arikan}. However, we have to point out that the SC's performance is seriously impacted by the degree of channel polarization. For short to medium code length, which is currently the main application of polar codes in 5G, virtual sub-channels are far from perfectly polarized \cite{Arikan}, so the SC may performs poorly. To achieve better performance, one may replace the SC with the BP \cite{BP}. 

\begin{figure*}[!t]
	\small
	\centering
	\setlength{\unitlength}{1cm}	
	\begin{tikzpicture}
	
	\foreach \y in {1,3,5,7} {
		\node[circle,draw]at(0, 12-1.5*\y){+};
		\node[circle,draw]at(0, 10.5-1.5*\y){=};
		\draw (-0.5,-2+1.5*\y)rectangle(0.5,0.5+1.5*\y); 
	}

	\foreach \y in {1,2,5,6} {
		\node[circle,draw]at(3, 12-1.5*\y){+};   
		\node[circle,draw]at(3, 9-1.5*\y){=};
	}
	\draw (3-0.5,-0.5)rectangle(3+0.5,5);
	\draw (3-0.5,+5.5)rectangle(3+0.5,11);
	
	\foreach \y in {1,2,3,4} {
		\node[circle,draw]at(6, 12-1.5*\y){+};   
		\node[circle,draw]at(6, 6-1.5*\y){=}; 
	}  
	\draw (6-0.5,-0.5)rectangle(6+0.5,11);

	\foreach \y in {1,...,8} {
		\node[draw,circle]at(-2.85, 12-1.5*\y){$u_{\y}$};
	}	

	\foreach \y in {1,...,8} {
		\node[draw,circle]at(8.85, 12-1.5*\y){$x_{\y}$};
	}	
	
	\foreach \y in {1,2} {
		\draw[red,<-] (-2.5, 12.1-1.5*\y)--(-0.5, 12.1-1.5*\y) node[midway,above]{$L_{3,\y}$};
	}
	\foreach \y in {1,2} {
		\draw[red,<-] ( 0.5, 12.1-1.5*\y)--( 2.5, 12.1-1.5*\y) node[midway,above]{$L_{2,\y}$};
	}
	\foreach \y in {1,...,4} {
		\draw[red,<-] ( 3.5, 12.1-1.5*\y)--( 5.5, 12.1-1.5*\y) node[midway,above]{$L_{1,\y}$};
	}
	\foreach \y in {1,2} {
		\draw[red,->,dashed] (0.5, 11.9-1.5*\y)--(2.5, 11.9-1.5*\y) 	node[midway,below]{$R_{2,\y}$};
	}	

	\foreach \y in {3,4} {
		\draw[cyan,<-] (-2.5, 12.1-1.5*\y)--(-0.5, 12.1-1.5*\y) 	node[midway,above]{$L_{3,\y}$};
	}
	\foreach \y in {3,4} {
		\draw[cyan,<-] ( 0.5, 12.1-1.5*\y)--( 2.5, 12.1-1.5*\y) node[midway,above]{$L_{2,\y}$};
	}
	\foreach \y in {3,4} {
		\draw[cyan,->,dashed] (0.5, 11.9-1.5*\y)--(2.5, 11.9-1.5*\y) 	node[midway,below]{$R_{2,\y}$};
	}
	\foreach \y in {1,2,3,4} {
		\draw[cyan,->,dashed] (3.5, 11.9-1.5*\y)--(5.5, 11.9-1.5*\y)  node[midway,below]{$R_{1,\y}$};
	}	

	\foreach \y in {5,...,8} {
	\draw[green,<-] ( 3.5, 12.1-1.5*\y)--( 5.5, 12.1-1.5*\y) node[midway,above]{$L_{1,\y}$};
}
	\foreach \y in {5,6} {
		\draw[green,<-] (-2.5, 12.1-1.5*\y)--(-0.5, 12.1-1.5*\y) node[midway,above]{$L_{3,\y}$};
	}
	\foreach \y in {5,6} {
		\draw[green,<-] (0.5, 12.1-1.5*\y)--(2.5, 12.1-1.5*\y) node[midway,above]{$L_{2,\y}$};
	}
	\foreach \y in {5,6} {
		\draw[green,->,dashed] (0.5, 11.9-1.5*\y)--(2.5, 11.9-1.5*\y) node[midway,below]{$R_{2,\y}$};
	}	

	\foreach \y in {7,8} {
		\draw[blue,<-] (0.5, 12.1-1.5*\y)--(2.5, 12.1-1.5*\y) node[midway,above]{$L_{2,\y}$};
	}
	\foreach \y in {7,8} {
		\draw[blue,<-] (-2.5, 12.1-1.5*\y)--(-0.5, 12.1-1.5*\y) node[midway,above]{$L_{3,\y}$};
	}
	\foreach \y in {7,8} {
		\draw[blue,->,dashed] ( 0.5, 11.9-1.5*\y)--( 2.5, 11.9-1.5*\y) node[midway,below]{$R_{2,\y}$};
	}

	\foreach \y in {5,6,7,8} {
		\draw[blue,->,dashed] (3.5, 11.9-1.5*\y)--(5.5, 11.9-1.5*\y)  node[midway,below]{$R_{1,\y}$};
	}		
	\foreach \y in {1,...,8} {
		\draw[->,dashed] (-2.5, 11.9-1.5*\y)--(-0.5, 11.9-1.5*\y) 	node[midway,below]{$R_{3,\y}$};
		\draw[<-] (6.5, 12.1-1.5*\y)--(8.5, 12.1-1.5*\y)  	node[midway,above]{$L_{0,\y}$};
		\draw[blue,->,dashed] (6.5, 11.9-1.5*\y)--(8.5, 11.9-1.5*\y)  node[midway,below]{$R_{0,\y}$};
	}		
	\end{tikzpicture}	
	\caption{An example of SCAN decoding for polar codes, where $L_{l,i}$'s refer to the LRs propagated from $x$-nodes to $u$-nodes, and $R_{l,i}$'s refer to the LRs propagated from $u$-nodes to $x$-nodes.}
	\label{fig:scan}
\end{figure*}
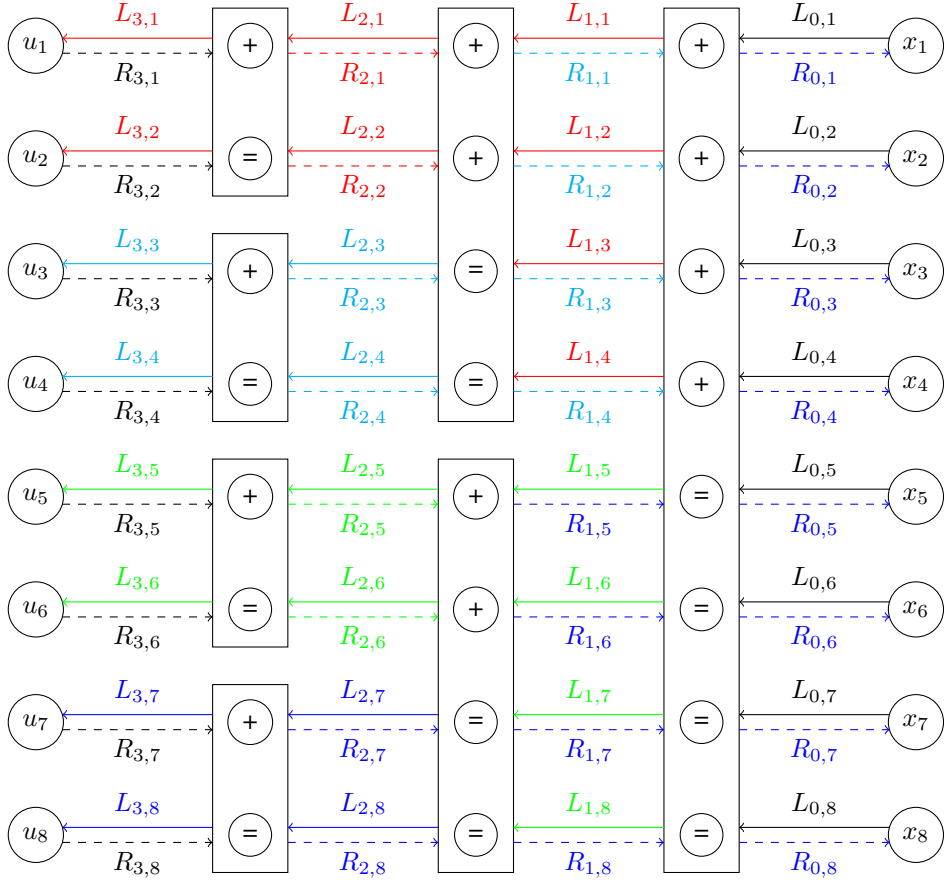

As shown by Fig.~\ref{fig:scan}, to realize the BP decoding for polar codes, we define two $(n+1)\times N$ matrices $(L_{l,i})_{(n+1)\times N}$ and $(R_{l,i})_{(n+1)\times N}$, where the former stores the Likelihood Ratios (LRs) propagated from $x$-nodes to $u$-nodes, while the latter stores the LRs propagated from $u$-nodes to $x$-nodes. For $u_i$, $R_{n,i}$ is the intrinsic LR and $L_{n,i}$ is the extrinsic LR; while for $x_i$, $L_{0,i}$ is the intrinsic LR and $R_{0,i}$ is the extrinsic LR. Initially, $L_{0,i} = \frac{W(y_i|0,s_i)}{W(y_i|1,s_i)}$ for all $i\in\{1,\cdots,N\}$, where $W(y_i|x_i,s_i)$ is the $i$-th physical sub-channel, and $R_{n,i}=+\infty$ for all $i\in\mathcal{A}_c$ (frozen bits are set to 0). For other $l$ and $i$, we set $L_{l,i}=R_{l,i}=1$. Then the BP decoding can be iterated.

Originally, the BP decoding for polar codes is very time-consuming \cite{BP}, so a fast variant---SCAN decoding---was proposed in \cite{SCAN}. The SCAN decoding includes one or more {\bf iterations}. At each iteration, $L_{n,i}$'s, the extrinsic LRs of $u$-nodes, and $R_{0,i}$'s, the extrinsic LRs of $x$-nodes, are calculated according to $L_{0,i}$'s, the intrinsic LRs of $x$-nodes, and $R_{n,i}$'s, the intrinsic LRs of $u$-nodes. Each SCAN iteration contains $N/2$ {\bf rounds} and further each round includes one $x$-to-$u$ {\bf LR-propagation} followed by one $u$-to-$x$ {\bf LR-propagation}. The best way to understand the SCAN algorithm is by an example. In Fig.~\ref{fig:scan}, the {\bf solid} lines with {\bf left} arrows form the $x$-to-$u$ LR flow, and the {\bf dashed} lines with {\bf right} arrows form the $u$-to-$x$ LR flow. The {\bf black} solid/dashed lines with left/right arrows denote the {\bf intrinsic} LRs of $x$-nodes/$u$-nodes. For $N=8$, each SCAN iteration includes four rounds, marked with different colors, \textcolor{red}{red} for the \textcolor{red}{first}, \textcolor{cyan}{cyan} for the \textcolor{cyan}{second}, \textcolor{green}{green} for the \textcolor{green}{third}, and \textcolor{blue}{blue} for the \textcolor{blue}{fourth}.

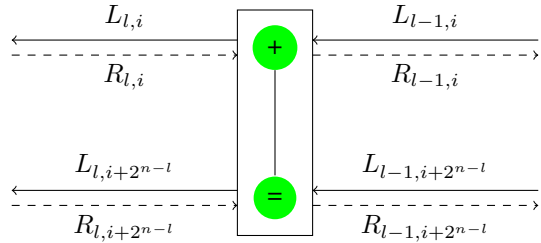
\begin{figure*}[!t]
	\small
	\centering
	\setlength{\unitlength}{1cm}
	
	\begin{tikzpicture}
	\node[circle,fill=green]at(0, 2){+};
	\node[circle,fill=green]at(0, 0){=};
	\draw (-0.5,-0.5) rectangle(0.5,2.5);		
	\draw (0,0.3)--(0,2-0.3);
	
	\draw[<-] (-4+0.5, +0.1  )--( -0.5, +0.1  ) node[midway,above]{$L_{l,i+2^{n-l}}$};
	\draw[<-] (-4+0.5, +0.1+2)--( -0.5, +0.1+2) node[midway,above]{$L_{l,i}$};
	
	\draw[<-] (   0.5, +0.1  )--(4-0.5, +0.1  ) node[midway,above]{$L_{l-1,i+2^{n-l}}$};
	\draw[<-] (   0.5, +0.1+2)--(4-0.5, +0.1+2) node[midway,above]{$L_{l-1,i}$};
	
	\draw[->,dashed] (-4+0.5, -0.1  )--( -0.5, -0.1  ) node[midway,below]{$R_{l,i+2^{n-l}}$};
	\draw[->,dashed] (-4+0.5, -0.1+2)--( -0.5, -0.1+2) node[midway,below]{$R_{l,i}$};
	
	\draw[->,dashed] (   0.5, -0.1  )--(4-0.5, -0.1  ) node[midway,below]{$R_{l-1,i+2^{n-l}}$};
	\draw[->,dashed] (   0.5, -0.1+2)--(4-0.5, -0.1+2) node[midway,below]{$R_{l-1,i}$};	
	\end{tikzpicture}
	
	\caption{LR-propagation kernel for polar decoding, where $1\leq l\leq n$ and $1\leq (i-k2^{n-l+1})\leq 2^{n-l}$ for $0 \leq k < 2^{l-1}$ .}
	\label{fig:kernel}
\end{figure*}

The $x$-to-$u$ LR-propagation calculates $L_{n,i}$'s by $n$ recursions, and the $u$-to-$x$ LR-propagation calculates $R_{0,i}$'s by $n$ recursions. Fig.~\ref{fig:kernel} shows the LR-propagation kernel. Let us define $a*b \triangleq \frac{ab+1}{a+b}$. For $1\leq (i-k2^{n-l+1})\leq 2^{n-l}$, where $1\leq l\leq n$ and $0 \leq k < 2^{l-1}$, the kernel for the $x$-to-$u$ LR-propagation is
\begin{align}
	\left\{
	\begin{array}{rl}		
		L_{l,i} 		&= L_{l-1,i} * (L_{l-1,i+2^{n-l}} \cdot R_{l,i+2^{n-l}})\\
		L_{l,i+2^{n-l}} &= (L_{l-1,i} * R_{l,i}) \cdot L_{l-1,i+2^{n-l}}
	\end{array},
	\right.
\end{align}
and the kernel for the $u$-to-$x$ LR-propagation is
\begin{align}
	\left\{
	\begin{array}{rl}		
		R_{l-1,i} &= R_{l,i} * (L_{l-1,i+2^{n-l}} \cdot R_{l,i+2^{n-l}})\\
		R_{l-1,i+2^{n-l}} &= (L_{l-1,i} * R_{l,i}) \cdot R_{l,i+2^{n-l}}
	\end{array}.
	\right.
\end{align}

After each SCAN iteration, the overall LR of $u_i$ is $(L_{n,i}\cdot R_{n,i})$. Then a hard decision is made to estimate $u_i$. Let $\hat{u}_i$ be the estimate of $u_i$. For $i\in\mathcal{A}$, $\hat{u}_i=\mathbf{1}_{[0,1)}{(L_{n,i}\cdot R_{n,i})}$, where
\begin{align}
\mathbf{1}_{\mathcal{I}}(x) \triangleq \left\{
\begin{array}{ll}
1, & x\in \mathcal{I}\\
0, & x\notin \mathcal{I}
\end{array}.
\right.
\end{align}
To verify the correctness of $\hat{u}^N$, one can send the CRC checksum \cite{crc} or Hash digest of $u^N$ together with $x^N$. After each SCAN iteration, if the CRC checksum or Hash digest of $\hat{u}^N$ matches the CRC checksum or Hash digest of $u^N$, the decoding is terminated; otherwise, one more iteration is run. For more detail about SCAN decoding, please refer to \cite{SCAN}.

\subsection{Decoding Correctness Verification}
Compared with LDPC codes, a defect of polar codes is that there is no explicit syndrome for decoding correctness verification. Though this problem can be solved by sending the CRC checksum \cite{crc} or Hash digest of $u^N$ together with $x^N$, it is more preferable if the decoding correctness of polar codes can be verified in the absence of CRC checksums or Hash digests. Fortunately, we find a very simple solution to this problem. After each SCAN iteration, besides $\hat{u}_i$, we also make a hard decision on $(L_{0,i}\cdot R_{0,i})$, the overall LR of $x_i$, to get $\hat{x}_i=\mathbf{1}_{[0,1)}{(L_{0,i}\cdot R_{0,i})}$. If $\hat{u}^NG_N=\hat{x}^N$, the decoding is terminated; otherwise, one more iteration is run. Now it can be seen that decoding correctness can be verified by polar codes themselves. As LDPC codes, to avoid endless loop, a threshold must be set as the maximum iteration number. If the iteration number is greater than the threshold, the decoding will be forcedly terminated.

\section{Channel Estimation}\label{sec:w2scan}
Now we focus on $L_{0,i}$'s, the intrinsic LRs of $x$-nodes. To trigger the BP or SCAN decoding, $L_{0,i}$'s must be seeded according to the local states of physical sub-channels. For conciseness, only the AWGN channel model will be handled below, while the deduction can be easily extended to other channel models. For the AWGN channel model, $y_i = (1-2x_i) + z_i$ is received at the decoder, where $z_i$ is the Gaussian noise of the $i$-th physical sub-channel. We use $\sigma_i^2$ to denote the variance of $z_i$. Then ideally, i.e., $\sigma^2_i$ is perfectly known at the decoder, $L_{0,i} = \exp(2y_i/\sigma_i^2)$. However, if the channel is time-varying, it is hardly possible to know the varying local states of sub-channels beforehand. Hence, we have to set $L_{0,i} = \exp(2y_i/\hat{\sigma}_i^2)$, where $\hat{\sigma}_i^2$ is the estimate of $\sigma_i^2$. Usually, $\hat{\sigma}_i^2 = \bar{\sigma}^2 \triangleq \frac{1}{N}\sum_{i=1}^{N}\sigma_i^2$ for all $i\in\{1,\cdots,N\}$, which is equivalent to taking the time-varying channel with local states $\sigma_i^2$'s as a stationary channel with global state $\bar{\sigma}^2$. 

Once the intrinsic LRs $L_{0,i}$'s are initialized, they will remain unchanged during the decoding. However, as analyzed in Sect.~\ref{sec:model}, due to the concavity of entropy, there will be a rate loss if the time-varying channel is treated as a stationary channel. For LDPC codes, if the intrinsic LRs of variable nodes are coarsely seeded, a significant gain can be achieved by elaborately refining the intrinsic LRs of variable nodes after each BP iteration. There are many schemes that can reach this goal. Among them, the SWBP may be the best one \cite{FangTCOM12,FangTCOM13} due to its high efficiency, low complexity, and strong robustness. 

Since the SCAN is a special kind of BP algorithm, naturally we suppose that it should be possible to extend the SWBP from LDPC codes to polar codes based on the platform of SCAN. We refer to such a scheme as Sliding-Window SCAN (SWSCAN). Further, we improve the SWSCAN by considering a sliding window with unequal tap weights. Such an enhancement of SWSCAN is referred to as Weighted-Window SCAN (W$^2$SCAN).

\subsection{Problem Formulation}\label{sec:full}
To avoid confusion, we will use $x_{1:N} \triangleq (x_1,\cdots,x_N)$ instead of $x^N$ in this section. In addition, for ease of presentation, we use $x_{\pi^{-1}(1:N)} \triangleq (x_{\pi^{-1}(1)},\cdots,x_{\pi^{-1}(N)})$ instead of $x_{1:N}$ to denote the output of polar encoder. After permutation, $x_{\pi(\pi^{-1}(1:N))} = x_{1:N}$ instead of $x_{\pi(1:N)}$ will be conveyed over slowly-varying physical channel $W_{1:N}$. At the decoder, $y_{1:N}=((1-2x_{1:N})+z_{1:N})$ is received, where $z_i$ is a realization of a zero-mean Gaussian random variable with variance $\sigma_i^2$. We can model $z_i^2$ as the sum of {\bf slowly-varying} $\sigma_i^2$ and a {\bf fast-varying} noise $\epsilon_i$: $z_i^2=\sigma_i^2+\epsilon_i$. After each SCAN iteration, we get $(L_{0,i}\cdot R_{0,i})$, the overall LR of $x_i$, which can be used to calculate the bias probability of $x_i$ as $p_i = \frac{1}{1+(L_{0,i}\cdot R_{0,i})}$. From $y_i$ and $p_i$, we can estimate $z_i^2$ as
\begin{align}
	\hat{z}^2_i \triangleq p_i(y_i+1)^2 + (1-p_i)(y_i-1)^2.
\end{align}
We write $\hat{z}_i^2 = z_i^2+\delta_i = \sigma^2_i + (\epsilon_i + \delta_i)$, where $\delta_i$ will converge to 0 as the decoding proceeds, if the decoding succeeds finally. Our problem is how to estimate $\sigma_{1:N}^2 \triangleq (\sigma_1^2,\cdots,\sigma_N^2)$ from $\hat{z}_{1:N}^2 \triangleq (\hat{z}_1^2,\cdots,\hat{z}_N^2)$. It can be seen that $\hat{z}_i^2$ is the sum of a {\bf slowly-varying} target signal ($\sigma_i^2$) and an additive {\bf fast-varying} {\bf zero-mean} noise ($\epsilon_i + \delta_i$), so the essence of this problem is actually low-pass filtering. There are many well-known methods for adaptive filter design, e.g., Wiener filter, Kalman filter, etc. However, these methods usually assume that the target slowly-varying signal and the additive fast-varying noise are stationary random processes with known spectral characteristics or known autocorrelation and cross-correlation, and they are separable in frequency domain. For $\hat{z}_i^2$, the spectral characteristics of $\sigma_i^2$ is unknown in advance, and $\delta_i$ is volatile as the decoding proceeds. Hence, these well-known methods do not work, and we must resort to a different solution.

\subsection{Sliding-Window SCAN}\label{sec:full}
By the {\bf conditionally-memoryless} assumption defined in Sect.~\ref{sec:model}, it can be deduced that $\epsilon_{1:N}$ are conditionally-memoryless given $\sigma_{1:N}^2$. Notice that $\delta_i$ is deduced from $R_{0,i}$, so it is obvious that the property of $\delta_{1:N}$ depends on code structure. Due to the regular structure of polar codes, adjacent elements of $\delta_{\pi^{-1}(1:N)}$ must be mutually dependent, even given  $\sigma_{1:N}^2$. However, after permutation, $\delta_{1:N}$ will be {\bf conditionally-memoryless} given $\sigma_{1:N}^2$. This is the second purpose of permutation (the first purpose is to accelerate the polarization of virtual sub-channels as shown in Sect.~III-A). Now $(\epsilon_{1:N} + \delta_{1:N})$ are {\bf conditionally-memoryless} and {\bf zero-mean} given $\sigma_{1:N}^2$. Another important assumption is that $\sigma_i^2$ is {\bf slowly-varying}, which implies that $\sigma_i^2$ can be estimated by a sequence of $\hat{z}_{i+i'}^2$'s belonging to a window centered at $\hat{z}_i^2$. To avoid out-of-bounds accesses, for window size $(2m+1)$, we pad $\hat{z}_{1:N}$ to get
\begin{align}
	\hat{z}_{(1-m):(N+m)} &\triangleq(\hat{z}_{1-m},\cdots,\hat{z}_0,\hat{z}_{1:N},\hat{z}_{N+1},\cdots,\hat{z}_{N+m})\nonumber\\
	&= (\hat{z}_{1+m},\cdots,\hat{z}_2,\hat{z}_{1:N},\hat{z}_{N-1},\cdots,\hat{z}_{N-m}),
\end{align}
i.e., $z_{i}=z_{2-i}$ for $(1-m)\leq i\leq 0$ and $z_i=z_{2N-i}$ for $(N+1)\leq i\leq (N+m)$. This is actually the symmetric padding. Then the simplest way to estimate $\sigma_i^2$ is
\begin{align}\label{eq:hat_sigma2}
	\hat{\sigma}^2_i = \frac{1}{2m+1}\sum_{i'=-m}^{m}\hat{z}^2_{i+i'}. 
\end{align}
After several simple operations, we can get
\begin{align}
	\hat{\sigma}^2_{i+1} = \hat{\sigma}^2_i + \frac{1}{2m+1}(\hat{z}^2_{i+1+m} - \hat{z}^2_{i-m}),
\end{align}
showing that $\hat{\sigma}_i^2$ can be deduced recursively by the {\bf sliding-window} style. Further, the order of complexity for calculating $\hat{\sigma}^2_{1:N}$ is $O(N)$, irrelevant to half window size $m$. For this reason, we refer to such a scheme as {\bf Sliding-Window SCAN} (SWSCAN). A similar scheme based on BP decoding of LDPC codes is called {\bf Sliding-Window BP} (SWBP) in \cite{FangTCOM12,FangTCOM13}. Note that there is no permutation for SWBP \cite{FangTCOM12,FangTCOM13}, because LDPC codes are a class of random codes.

As shown by \eqref{eq:hat_sigma2}, the most important thing for the SWSCAN is how to set half window size $m$. Once again, to find the optimal half window size, we need the aid of the {\bf conditionally-memoryless} assumption. Let us slightly modify \eqref{eq:hat_sigma2} to 
\begin{align}\label{eq:hat_sigma2m}
	\hat{\sigma}^2_i(m) = \frac{1}{2m}\sum_{i'=1}^{m}(\hat{z}^2_{i-i'}+\hat{z}^2_{i+i'}).
\end{align}
By comparing \eqref{eq:hat_sigma2m} with \eqref{eq:hat_sigma2}, the reader can find that $\hat{z}_i^2$ is excluded from the estimate of $\sigma_i^2$, which makes $\hat{\sigma}_i^2(m)$ and $\hat{z}_i^2$ mutually {\bf conditionally independent} given $\sigma_i^2$. In other words, $\hat{\sigma}_i^2(m)$ and $\hat{z}_i^2$ can be taken as two {\bf independent observations} of $\sigma_i^2$, i.e., $\hat{z}_i^2\leftrightarrow\sigma_i^2\leftrightarrow\hat{\sigma}_i^2(m)$. Let $E(m)\triangleq\frac{1}{N}\sum_{i=1}^{N}e_i^2$, where $e_i\triangleq\hat{\sigma}_i^2(m)-\hat{z}_i^2$. Then the optimal $m$ in the sense of Mean-Squared-Error (MSE) is
\begin{align}\label{eq:dotm}
	\dot{m} = \arg\min_{m}E(m).
\end{align}
From \eqref{eq:hat_sigma2m}, it is easy to get
\begin{align}
e_{i+1} = e_i + \frac{1}{2m}(\hat{z}_{i+1+m}^2-\hat{z}_{i-m}^2) + (1+\frac{1}{2m})(\hat{z}_i^2-\hat{z}_{i+1}^2).
\end{align}
Hence, $e_i$ can be calculated recursively by a sliding window, and further, the order of complexity for calculating $E(m)$ is $O(N)$. In view of the low complexity of $E(m)$, we suggest implementing \eqref{eq:dotm} simply by a full search over $m\in\{1,\cdots,\lfloor N/2\rfloor\}$. If so, the order of complexity of SWSCAN is $O(N^2)$, the same as that of SWBP \cite{FangTCOM12,FangTCOM13}.

\subsection{Weighted-Window SCAN}\label{sec:frequency}
As shown by \eqref{eq:hat_sigma2m}, the SWSCAN (and SWBP for LDPC codes) is actually equivalent to a low-pass filter with {\bf equal tap weights} $1/(2m)$. To this point, the reader may immediately come up with the idea of {\bf unequal tap weights}. That is just what we will discuss below.

Let $(w_{-m},\cdots,w_{-1},w_0,w_1,\cdots,w_m)$ be the tap weight vector, where $w_0\equiv 0$. For simplicity, it is very natural to use symmetric taps, i.e., $w_{i'}=w_{-i'}$ for $1\leq i'\leq m$. Let $a_{i+i'} \triangleq \hat{z}^2_{i-i'}+\hat{z}^2_{i+i'}$ and ${\bf a}_i \triangleq (a_{i+1},\cdots,a_{i+m})^\top\in \mathbb{R}^{m\times 1}$ for $1\leq i\leq N$. Let ${\bf w} \triangleq (w_1,\cdots,w_m)^\top\in \mathbb{R}^{m\times 1}$. Then
\begin{align}
	\hat{\sigma}^2_i({\bf w}) = \sum_{i'=1}^{m}w_{i'}(\hat{z}^2_{i-i'}+\hat{z}^2_{i+i'}) = {\bf a}_i^\top {\bf w}.
\end{align}
It is easy to get
\begin{align}
	(\hat{\sigma}_i^2({\bf w})-\hat{z}_i^2)^2 
	&= {\bf w}^\top {\bf a}_i{\bf a}_i^\top {\bf w} - 2\hat{z}^2_i{\bf a}_i^\top{\bf w} + \hat{z}^4_i\nonumber\\
	&= {\bf w}^\top {\bf H}_i {\bf w} - 2{\bf f}_i^\top {\bf w} + \hat{z}^4_i,
\end{align}
where ${\bf H}_i = {\bf a}_i{\bf a}_i^\top\in\mathbb{R}^{m\times m}$ and ${\bf f}_i = \hat{z}^2_i{\bf a}_i \in \mathbb{R}^{m\times 1}$. Let ${\bf H} = \sum_{i=1}^{N}{\bf H}_i$ and ${\bf f} = \sum_{i=1}^{N}{\bf f}_i$. Then
\begin{align}
	E({\bf w}) \triangleq \sum_{i=1}^{N}(\hat{\sigma}^2_i({\bf w})-\hat{z}^2_i)^2 = {\bf w}^\top {\bf H} {\bf w} - 2{\bf f}^\top {\bf w} + \sum_{i=1}^{N}\hat{z}^4_i.
\end{align}
This is a standard least-square problem with solution
\begin{align}
	\dot{\bf w} = \arg\min_{\bf w}E({\bf w}) = {\bf H}^{-1}{\bf f}.
\end{align}
However, the optimal ${\bf w}$ obtained by the least-square method performs very poorly in practice. The reason is because that there is no constraint on ${\bf w}$. Considering its physical meanings, the following constraints on ${\bf w}$ are obvious:
\begin{itemize}
	\item {\bf Non-negativity}: $w_{i'}\geq 0$;
	\item {\bf Monotonicity}: $w_{i'}\geq w_{i'+1}$;
	\item {\bf Normality}: $\sum_{i'=1}^{m}2w_{i'}=1$.
\end{itemize}
Imposed by the above constraints, the problem is now formatted as
\begin{align}\label{eq:dotw}
	\dot{\bf w} = \arg\min_{\bf w}({\bf w}^\top {\bf H} {\bf w} - 2{\bf f}^\top {\bf w})\quad{\rm s.t.}\quad w_{i'}\geq 0, w_{i'}\geq w_{i'+1}, \sum_{i'=1}^{m}w_{i'}=0.5. 
\end{align}
This is a standard quadratic programming problem that can be solved effectively in (weakly) polynomial time. We refer to this scheme as Weighted-Window SCAN (W$^2$SCAN).

\subsection{Complexity of W$^2$SCAN}\label{subsec:complexity}
Now we consider how to calculate ${\bf H}$ and ${\bf f}$. Let ${\bf H} = (h_{k,l})_{m\times m}$ and ${\bf f} = (f_k)_{m\times 1}$, where $1\leq k,l\leq m$. Obviously,
\begin{align}
	h_{k,l} = h_{l,k} = \sum_{i=1}^{N}a_{i+k}a_{i+l} =  \sum_{i=1}^{N}(\hat{z}^2_{i-k}+\hat{z}^2_{i+k})(\hat{z}^2_{i-l}+\hat{z}^2_{i+l}).
\end{align}
and
\begin{align}
	f_k = \sum_{i=1}^{N}\hat{z}^2_ia_{i+k} = \sum_{i=1}^{N}\hat{z}^2_i(\hat{z}^2_{i-k}+\hat{z}^2_{i+k}).
\end{align}
Let ${\bf \Phi} \triangleq (\phi_{k,l})_{(2m+1)\times(2m+1)} \in \mathbb{R}^{(2m+1)\times(2m+1)}$, where $-m\leq k,l\leq m$, and
\begin{align}\label{eq:phi}
	\phi_{k,l} = \phi_{l,k} \triangleq \sum_{i=1}^{N}\hat{z}^2_{i+k}\hat{z}^2_{i+l}. 
\end{align}
It can be found from \eqref{eq:phi} that $\phi_{k,l}$ is similar to but slightly different from the autocorrelation of $\hat{z}^2_{1:N}$. Then it is easy to get $h_{k,l} = \phi_{-k,-l} + \phi_{-k,l} + \phi_{k,-l} + \phi_{k,l}$ and $f_k = \phi_{k,0} + \phi_{-k,0}$. Thus, if only ${\bf \Phi}$ is known, ${\bf H}$ and ${\bf f}$ can be easily calculated. 

According to the definition of ${\bf\Phi}$, the number of operations to calculate ${\bf \Phi}$ is proportional to $N(2m+1)^2$. So the order of computational complexity of ${\bf \Phi}$ is $O(Nm^2)$, which tells us that the half window size $m$ is a very important factor impacting the complexity of W$^2$SCAN. How to set $m$ is a tradeoff between efficiency and complexity: Larger $m$ will improve coding efficiency but increase computational complexity, and vice versa. 

To find an appropriate value for the half window size $m$, we propose to combine the W$^2$SCAN with the SWSCAN. After {\bf each} SCAN iteration, we first run \eqref{eq:dotm} to find the optimal half window size $\dot{m}$ for the SWSCAN with {\bf equal}-weight taps. Obviously, the optimal half window size for the W$^2$SCAN with {\bf unequal}-weight taps must be not smaller than $\dot{m}$. Hence, we set the half window size $m$ of W$^2$SCAN to a value not smaller than $\dot{m}$. After that, \eqref{eq:dotw} is run to find the optimal $m$ tap weights $\dot{\bf w}$. An interesting finding from experiments is that a half window size larger than $\dot{m}$ can bring only a negligible gain for the W$^2$SCAN (see the next section). Hence, we suggest $m=\dot{m}$, i.e., setting the half window size of W$^2$SCAN to $\dot{m}$, the optimal half window size of SWSCAN, which balances efficiency and complexity well.

\section{Experimental Results}\label{sec:exp}
The author has independently realized the encoder, the SC decoder, and the SCAN decoder of polar codes with MATLAB. Then on the platform of SCAN decoder, both SWSCAN and W$^2$SCAN decoders are implemented. For the W$^2$SCAN decoder, the quadratic programming is actualized with the $\rm{quadprog}$ function of MATLAB.

Since similar phenomena are observed under different settings, to avoid prolixity, only the results under the following settings are reported. The code length is $N=2^{10}=1024$ and the code rate is $R=0.5$, i.e., only $K=NR=512$ best virtual sub-channels are utilized for user bits. The physical channel is piecewise-stationary and the length of piece is Poisson-distributed with parameter $\lambda=64$. For each piece, the noise variance of the AWGN physical channel is uniformly drawn from the space $\mathcal{S} = \{0,\bar{\sigma}^2,2\bar{\sigma}^2\}$, where $\bar{\sigma}^2\in\{0.45,0.5,0.55,0.6,0.65,0.7\}$. If the reader is interested in related work, he or she can download and run the software package released in the author's homepage \cite{Fang} to get more results under other settings. 

We use the Monte Carlo simulation to construct good polar codes for the AWGN physical channel. Note that in practice, the codec is unaware of the non-stationarity of the channel before transmission, so we take the time-varying AWGN channel as a stationary AWGN channel with constant noise variance $\bar{\sigma}^2$ during code construction. We run $10^5$ trials for each $\bar{\sigma}^2$ and sort all virtual sub-channels according to their Bit-Error-Rates (BERs). Finally, the best $K=512$ virtual sub-channels are dedicated to user bits.

\begin{figure*}
	\subfigure[]{\includegraphics[width=.5\linewidth]{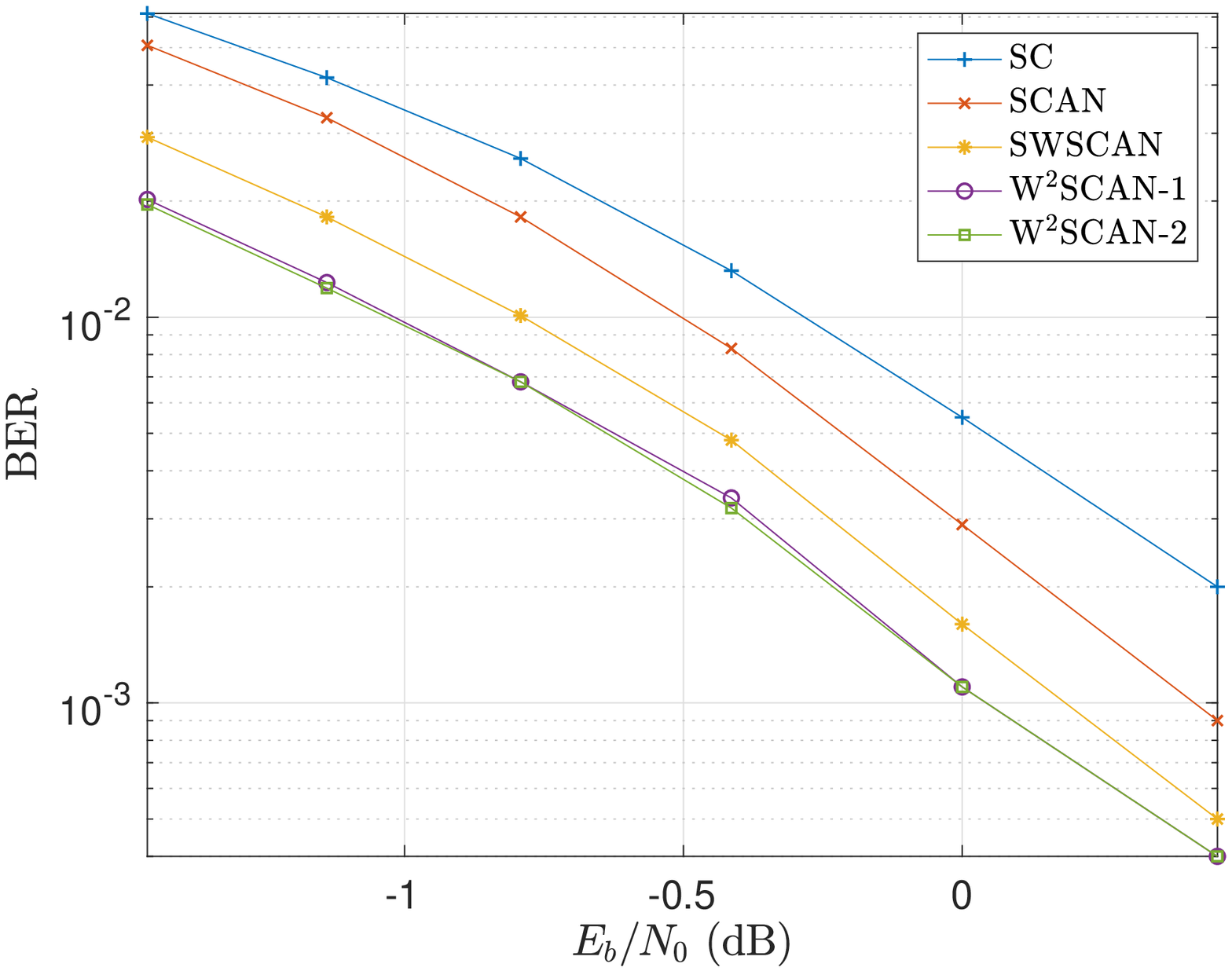}}%
	\subfigure[]{\includegraphics[width=.5\linewidth]{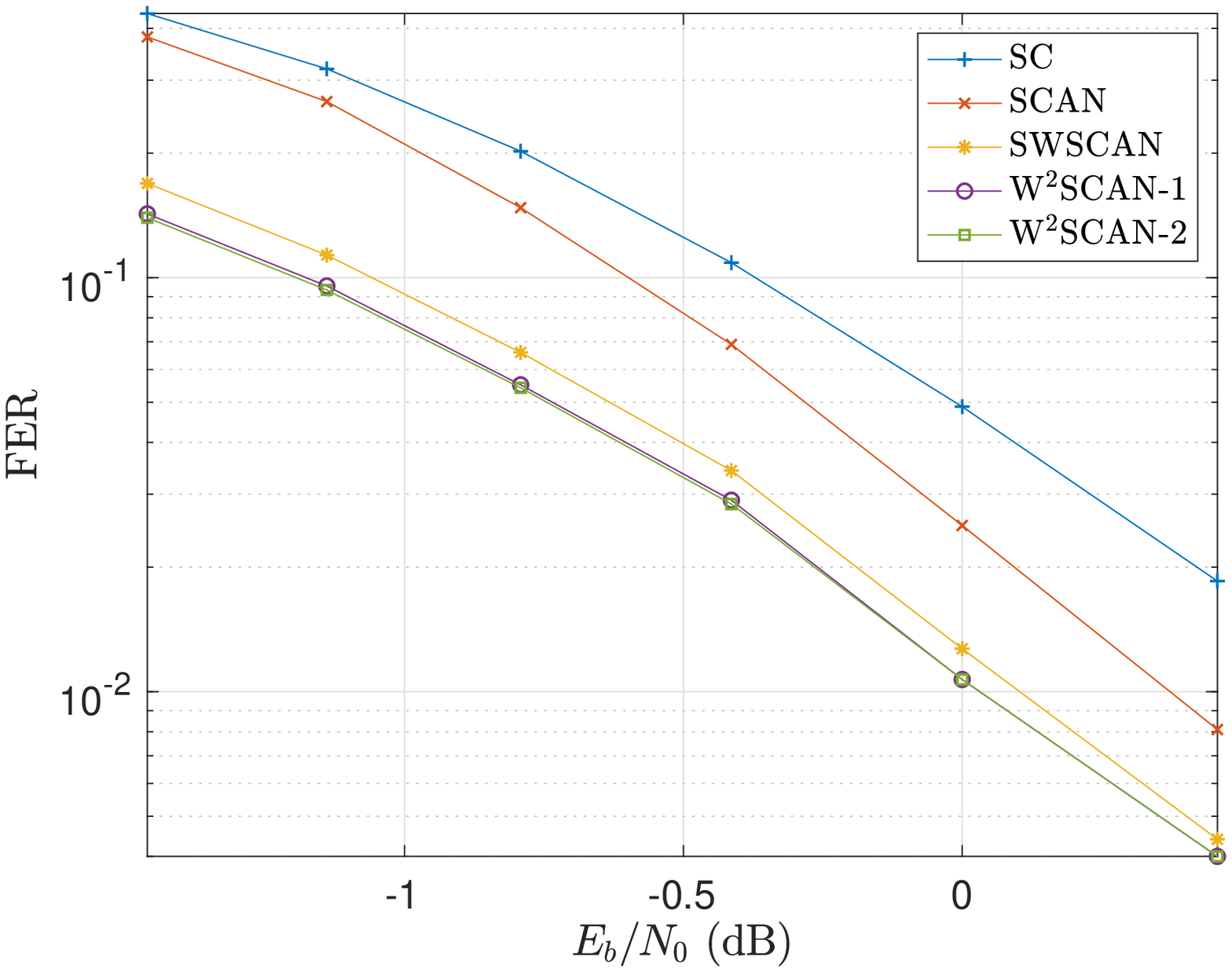}}
	\caption{Results of bit-error-rate (BER) and frame-error-rate (FER).}
	\label{fig:results}
\end{figure*}

After polar codes are constructed, different decoders are evaluated. First, the user bits $u_{\mathcal{A}}$ are uniformly generated and the forbidden bits $u_{\mathcal{A}^c}$ are set to 0. After encoding, $x^N$ is {\bf randomly permuted} before transmission. At the receiver, $\hat{\sigma}_i^2$, $\forall i\in\{1,\cdots,N\}$, is initialized to $\bar{\sigma}^2$, and then $y^N$ is decoded with different methods. For the SC and the SCAN, $\hat{\sigma}_i^2$ remains unchanged, while for the SWSCAN and the W$^2$SCAN, $\hat{\sigma}_i^2$ is updated after {\bf each} SCAN iteration. For the SCAN, the SWSCAN, and the W$^2$SCAN, at most $(n+1)$ iterations are attempted. We run $10^5$ trials for each $\bar{\sigma}^2$. The statistical results are included in Fig.~\ref{fig:results}, where W$^2$SCAN-$\alpha$ refers to the W$^2$SCAN with half window size $m=\alpha\dot{m}$, where $\dot{m}$ is defined by \eqref{eq:dotm}. Fig.~\ref{fig:results}(a) shows the BER versus $E_b/N_0$, and Fig.~\ref{fig:results}(b) shows the Frame-Error-Rate (FER) versus $E_b/N_0$, where $E_b/N_0 = -10\log_{10}(2\bar{\sigma}^2)$. It can be observed from Fig.~\ref{fig:results} that in order of SC, SCAN, SWSCAN, and W$^2$SCAN, both BER and FER are significantly lowered in turn. These results coincide with what reported in \cite{FangTCOM12,FangTCOM13} for the SWBP on the platform of LDPC codes. Another appealing finding from Fig.~\ref{fig:results} is that for the W$^2$SCAN, increasing half window size $m$ from $\dot{m}$ to $2\dot{m}$ can hardly bring any gain, so it is better to set $m$ to a relatively small value for low complexity.

\begin{figure}
	\subfigure[]{\includegraphics[width=.5\linewidth]{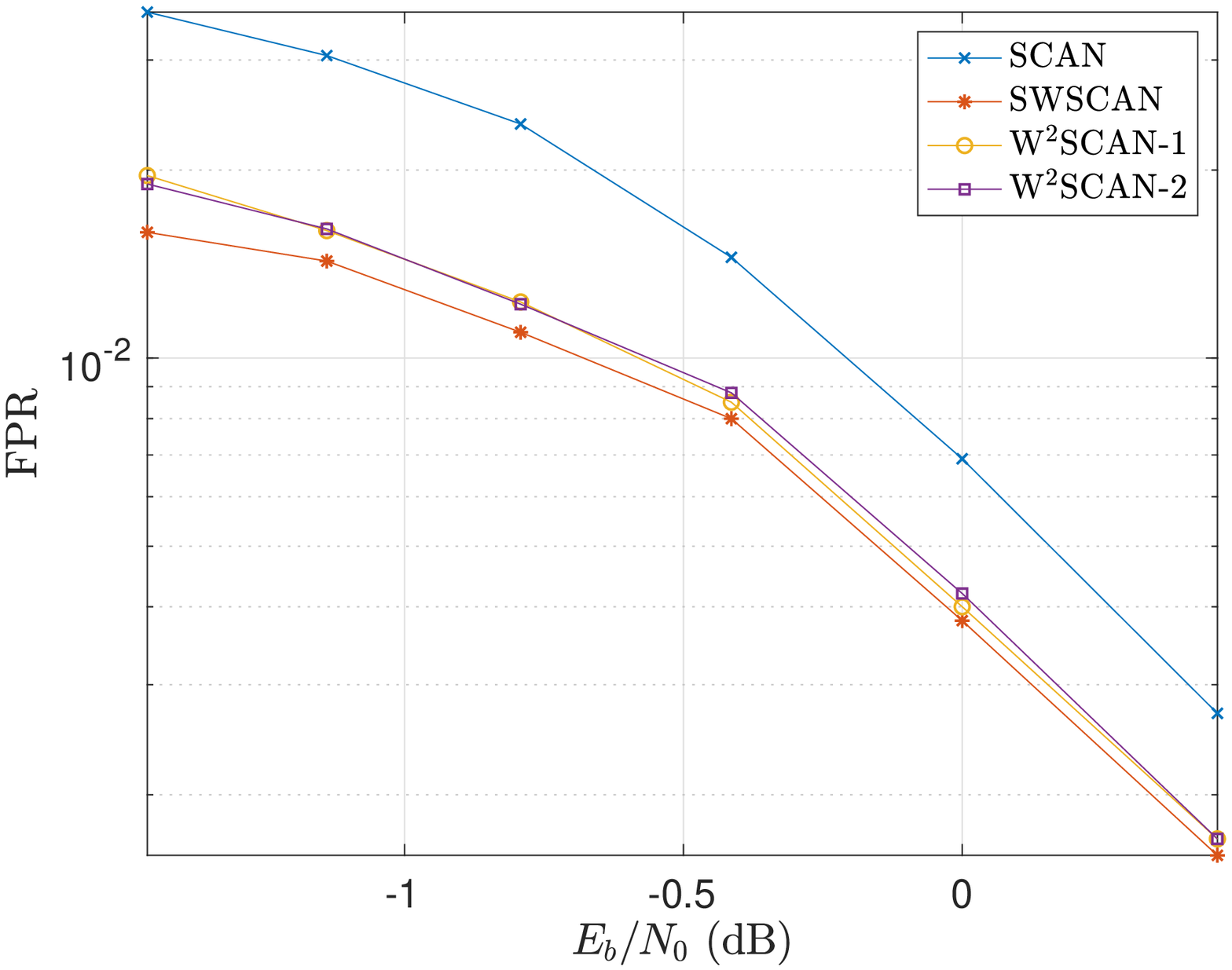}}%
	\subfigure[]{\includegraphics[width=.5\linewidth]{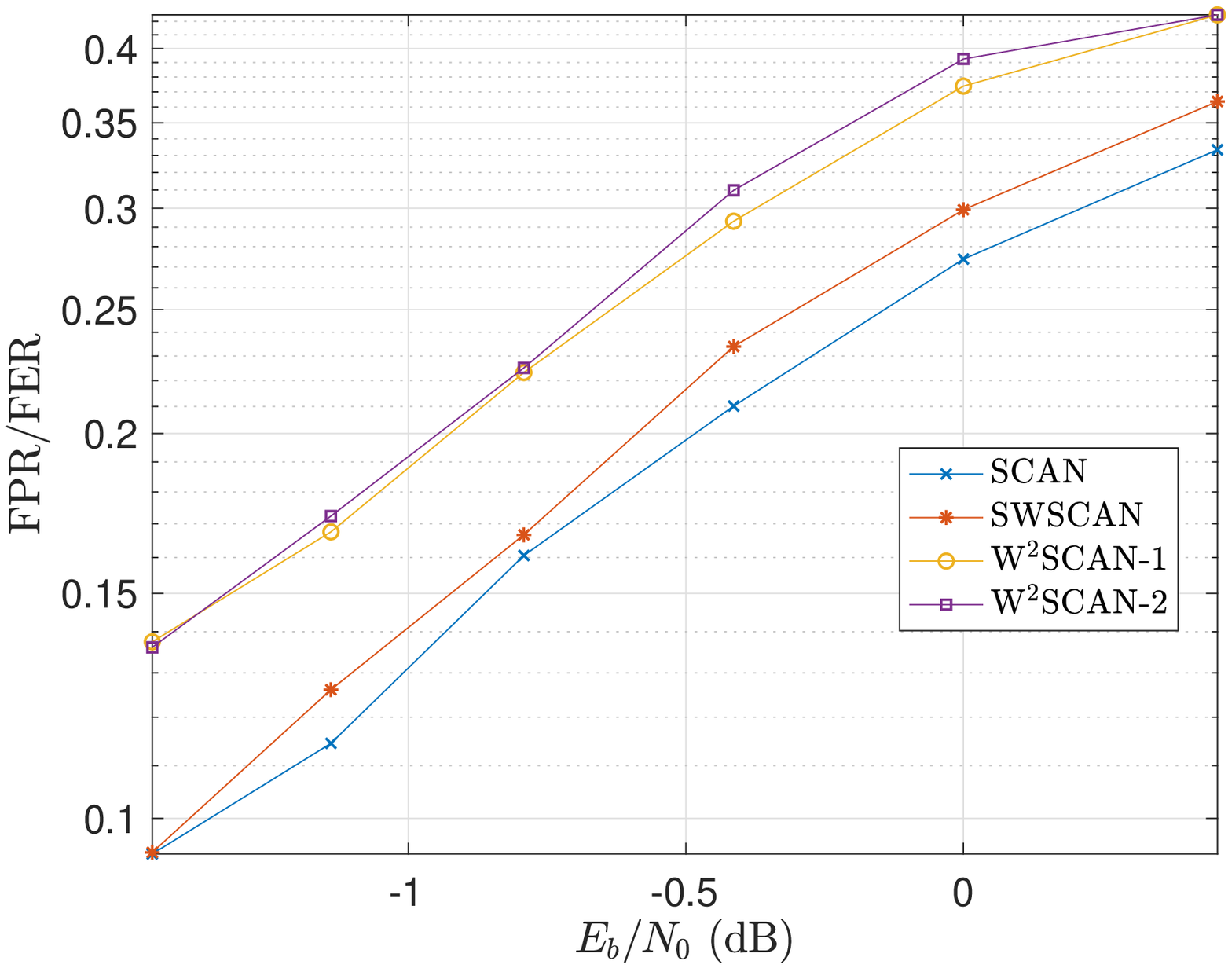}}
	\caption{Results of false-positivity-rate (FPR).}
	\label{fig:fpr}
\end{figure}

Just as BP decoding of LDPC codes, the SCAN and its variants may cause false positivity. To show this point, the False-Positivity-Rate (FPR) is plotted in Fig.~\ref{fig:fpr}(a), and the ratio of FPR to FER is plotted in Fig.~\ref{fig:fpr}(b). It can be seen that for the SCAN and its variants, as $E_b/N_0$ increases, the FPR descends but the ratio of FPR to FER ascends. That means: for high $E_b/N_0$, most decoding failures are caused by false positivity. Compared with the SCAN, its variants, including both SWSCAN and W$^2$SCAN, significantly lower the FPR, but heighten the ratio of FPR to FER. Finally, compared with the SWSCAN, the W$^2$SCAN increases decoding failures caused by false positivity, though it reduces the FER as a whole. 

As for running time, as expected, the SCAN is slower than the SC. However surprisingly, in some cases, the SWSCAN is even faster than the SCAN. We suppose that there are two reasons for this phenomenon. On one hand, \eqref{eq:dotm} is very simple, and on the other, fewer iterations are needed by the SWSCAN due to finer estimates of channel states. Similar phenomenon is also observed for the SWBP based on LDPC codes. Finally, the W$^2$SCAN is much slower than the SWSCAN as the quadratic programming is very time-consuming. We do not include the detailed results of running time in this paper because they heavily depend on how powerful the used computer is. If the reader is interested in this topic, he or she can run the program in \cite{Fang}.

\section{Conclusion}\label{sec:con}
This paper studies such a problem that the state of the physical channel is {\bf slowly-varying} and {\bf unknown} at the decoder. It is shown that, by {\bf permuting} codewords before transmission and adopting the {\bf SCAN} decoder, a similar idea can be borrowed from the SWBP, which is based on LDPC codes, and applied to polar codes. This scheme, named as SWSCAN, re-estimates channel states and updates the associated intrinsic LRs after each SCAN iteration during decoding. The SWSCAN can be further improved by introducing {\bf unequal} tap weights optimized with the {\bf quadratic programming}. This scheme is named as W$^2$SCAN. Experimental results show that the SWSCAN is significantly superior to the SCAN, and further the W$^2$SCAN is significantly superior to the SWSCAN.


\end{document}